\documentclass[twoside,a4paper,11pt]{proceedings}
\usepackage{graphicx}
\usepackage{hyperref}
\usepackage{movie15}
\usepackage{natbib}
\begin{document}
\pagenumbering{arabic}
\pagestyle{myheadings}
\thispagestyle{empty}
\vspace*{-1cm}
{\flushleft\includegraphics[width=3cm,viewport=0 -60 300 -20]{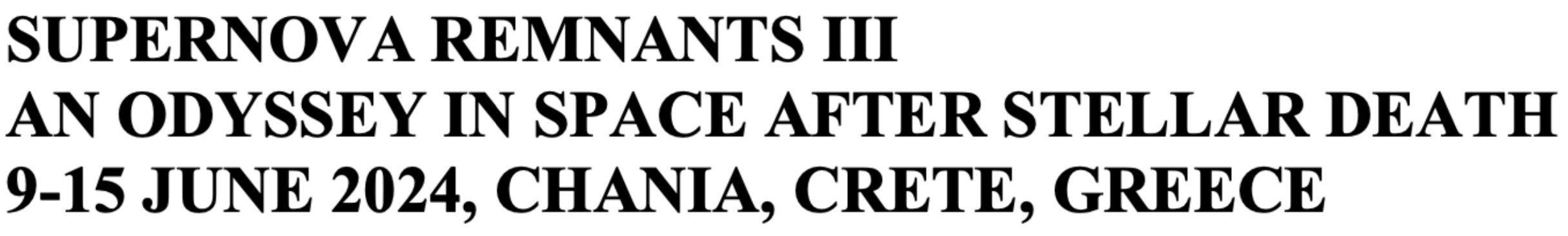}}
\vspace*{0.2cm}
\begin{flushleft}
{\bf {\LARGE
Characterization of M51's supernova remnants
with the imaging spectrometer SITELLE
}\\
\vspace*{0.5cm}
Billy Gamache,
Laurent Drissen,
Carmelle Robert
and Mykola Posternak
%
}\\
\vspace*{0.5cm}
%

Département de physique, de génie physique et d’optique, Université Laval, Québec (QC), G1V 0A6, Canada \\

%
\end{flushleft}
\markboth{
Supernova Remnants are the best
}{
Gamache et al.
}
\thispagestyle{empty}
\vspace*{0.4cm}
\begin{minipage}[l]{0.09\textwidth}
\ 
\end{minipage}
\begin{minipage}[r]{0.9\textwidth}
\section*{Abstract}{\small
We present preliminary results of a detailed 3D study of supernova remnants in the nearby spiral M51 using data from the SIGNALS survey obtained with the imaging Fourier transform spectrometer SITELLE at the Canada-France-Hawaii telescope (CFHT). Data cubes covering the entire galaxy were gathered in three spectral ranges: SN3 (647 - 685 nm, R = 5000), SN2 (482 - 513 nm, R = 600) and SN1 (363 - 386 nm, R = 1000). The spectral resolution of the SN3 cube allows a precise, spatially resolved measurement of the velocity dispersion of each object.  While most of the SNRs were known from previous surveys based on imagery and long-slit spectroscopy, we now provide 2D line flux and kinematic maps for all of them and found 20 new candidates. Most of the SNRs show velocity dispersions ($\sigma$) in the range $30-80$ km/s, which is typical for middle-aged SNRs. Finally, we compare the properties of SNRs with those of thousands of HII regions included in the same dataset.

\vspace{5mm}
\normalsize}
\end{minipage}

\section{Introduction}
Extragalactic supernova remnant (SNR) population studies are key elements for our understanding of the chemical evolution of galaxies. Each population has its own unique signature that reflects the star formation rate, metallicity and interstellar medium (ISM) properties of the host galaxy. The SNRs of M51, a well-studied, nearby ($\sim$ 8.6 Mpc) interacting system, was the subject of a recent study by \citet{winkler_optical_2021} (hereafter W21). Using Hubble Space Telescope images and Gemini Multi-Object Spectrograph (GMOS), they found 179 candidates and obtained the spectrum of 66 of them. Following their lead, we used data from the SIGNALS survey to characterize all of these candidates and search for new ones.

\section{Data and methodology}

M51's data are part of the SIGNALS survey\citep{Signals19}, wich targets nearby galaxies to study star formation through their emission lines with SITELLE \citep{Drissen2019}, an imaging Fourier transform spectrometer attached to the 3.6-m CFHT. This instrument produces data cubes with a $11'\times 11'$ field of view sampled at $0.32''$ per pixel; the spectral resolution can be tailored to the need of the program up to R = 10000. For this study, we used an $R = 5000$ cube with the SN3 filter (647 - 685 nm) for the detection process and spectral analysis. Its spectral range includes H$\alpha$, [NII]$\lambda\lambda6548,83$ and [SII]$\lambda\lambda6716,31$. Further analysis will be made with the SN2 (R = 600 for H$\beta$ and [OIII]$\lambda\lambda$4959, 5007) and SN1 (R = 1000 for [OII]$\lambda$3727) filters.

As a first step, we identified all of the W21 candidates in the emission line maps and searched for new objects having a [SII]$\lambda6716+\lambda6731$/H$\alpha$ ratio larger than $\sim$ 0.4 (the ``traditional'' criterion to detect extragalactic SNRs). This led to an initial list of 283 objects. For each of them, we selected a domain of integration using a flux treshold as well as a background region. Using the SITELLE dedicated software \texttt{ORCS} \citep{martin_data_2021}, we extracted the integrated, background-subtracted spectrum of each region and fitted each emission line with a sincgauss profile.\footnote{A sincgauss is the convolution of SITELLE's instrument line shape, a sinc function, with a gaussian representing the velocity dispersion along the line of sight.} \texttt{ORCS} then provides the amplitude, the flux, the velocity, the velocity dispersion (from the $\sigma$ of the gaussian) and the S/N ratio for each line. 



\begin{figure*}
    \centering
    \includegraphics[width=0.7\linewidth]{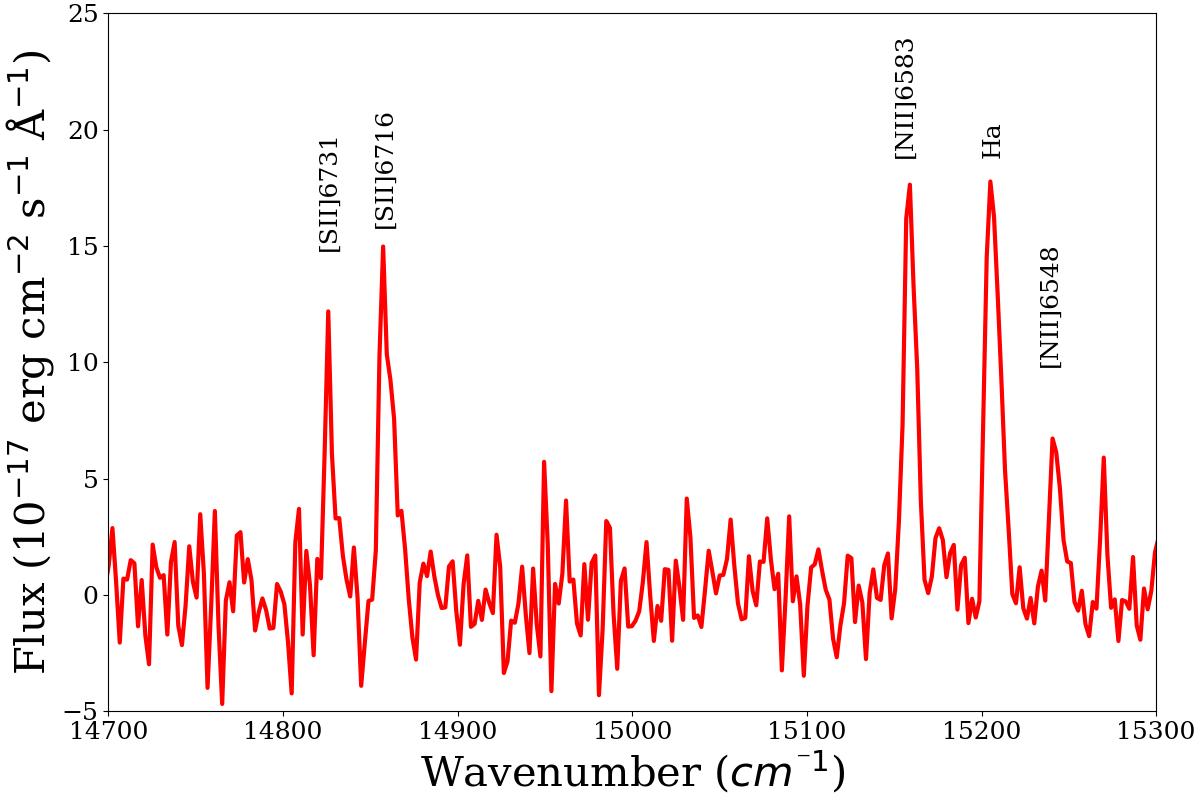}
    \caption{Spectrum of the newly found SNR candidate GD24-252 in the SN3 bandpass.}
    \label{spectre}
\end{figure*}

\section{Preliminary results and future work}
 As suggested by \cite{2018ApJ...855..140L} and \citet{2019ApJ...887...66P}, our team took advantage of the moderately high spectral resolution provided by the SN3 cubes of the SIGNALS survey to include the velocity dispersion as a supplementary criterion to identify SNRs: see for instance \citet{vicens-mouret_planetary_2023} and \citet{duarte_puertas_properties_2024}, who introduced the parameter $\xi =\frac{[SII]}{H\alpha} \times \sigma$. For our preliminary analysis of the M51 data, we thus chose to consider as excellent SNR candidates the objects for which the integrated spectrum simultaneously shows (a) an [SII]/H$\alpha$ ratio larger than 0.35, (b) a velocity dispersion of the [NII] lines, $\sigma_{[NII]}$, larger than 30 km~s$^{-1}$, as well as (c) a S/N ratio larger than 3 in all the lines. This led to a catalogue of 104 objects (84 from W21 and 20 new), which we considered for further analysis. Fig. \ref{spectre} shows the spectrum of one of these new SNR candidates.

\begin{figure}
    \centering
    \includegraphics[width=\linewidth]{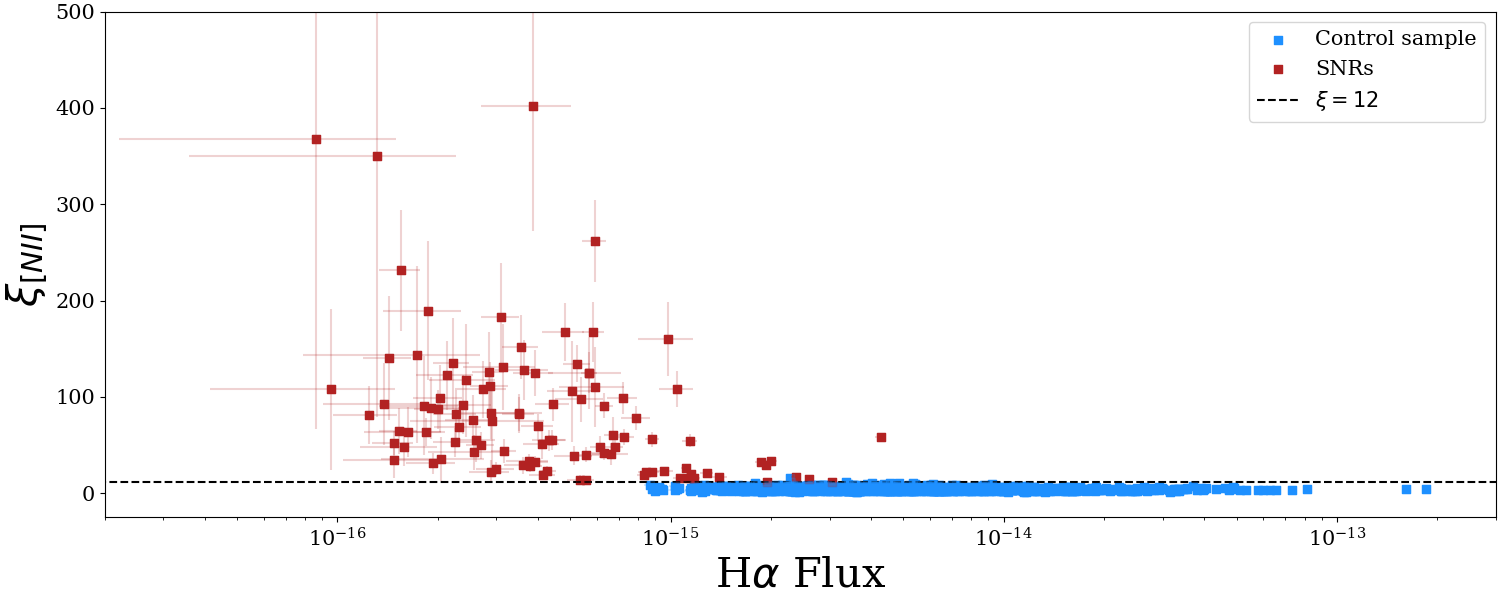}
    \caption{$\xi$ and H$\alpha$ flux for the candidate SNRs and the control sample in M51.}
    \label{xi_nii}
\end{figure}

For comparaison, we used a control sample of over 1000 emission regions (most likely HII regions) in the same data cube for which we also extracted the sky-subtracted integrated spectrum. Figure \ref{xi_nii} compares the SNR candidates population with the control sample in terms of the $\xi_{[NII]}$ parameter . The two populations are well separated, showing how SNRs are dominated by the shock heating process, but also by the kinematics of the gas.

\begin{figure}
    \centering
    \includegraphics[width=\linewidth]{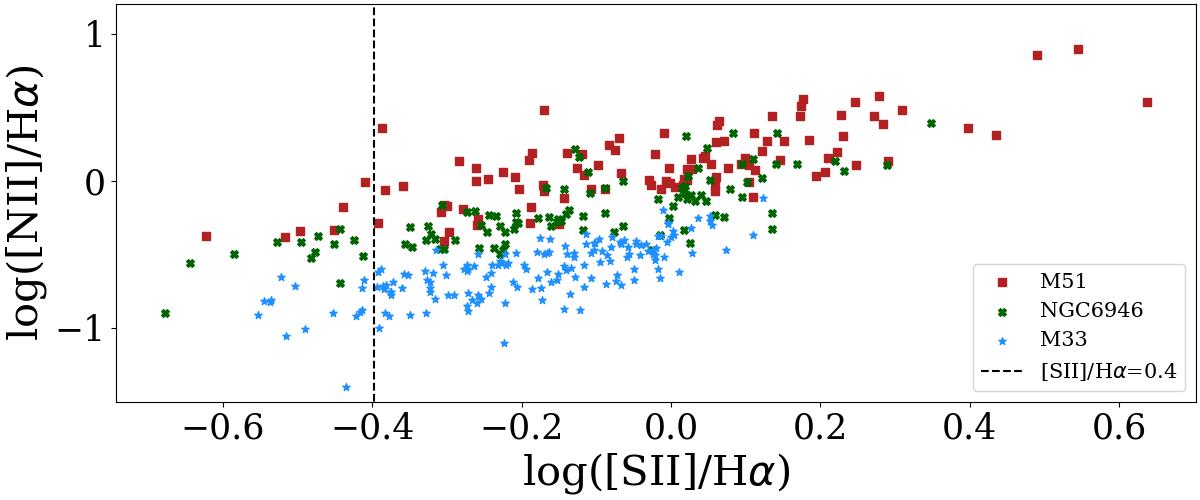}
    \caption{Line ratios for the SNRs in M51, in comparaison with those of NGC6946 \citep{long_new_2019} and M33 \citep{duarte_puertas_properties_2024}.}
    \label{spiral}
\end{figure}

The median value of [SII]/H$\alpha$ ratio is $1$ for SNR candidates, while the velocity dispersion of their [NII] lines ranges between $30$ and $160$ km/s. Their [NII]/H$\alpha$ ratio is also unusually high, with a median value of $1.6$. This property has already been noticed by \citet{winkler_optical_2021} and is associated with the high metallicity in the galaxy gas. This is shown in figure \ref{spiral} which compares M51 SNR line ratios to those of two other spirals. We can clearly see the odd nature of M51's SNRs. In addition to an strong average value, the [NII]/H$\alpha$ ratio of the SNR population also shows a clear negative galactocentric gradient, as expected. But interestingly, this is not the case of the HII regions, wich show a surprising positive gradient (excluding the central AGN and its surroundings).

The velocity dispersion measurements of the SNR population also show intriguing features. We sometimes found large differences between the velocity dispersion of the H$\alpha$ and the [NII] lines. For 33 cases, the difference is greater than $30$ km/s, H$\alpha$ always beeing the narrower line. Visual inspection of the spectra motivated us to model the spectra with two components for H$\alpha$ and [NII] in a few cases. We were able to fit 20 spectra with a broad and a narrow component with a S/N above 3 (5 with a S/N above 5). The velocity dispersion of the narrow component is in the range $10-25$ km/s while the broad one varies between 50 and 150 km/s. The broad component is tought to originate from the shock-heated plasma, but the narrow component is more mysterious. Despite our care in subtracting a proper local background, we cannot exclude the presence of some residuals. Another possibility is that it originates from a photoionised precursor \citep{medina_electron-ion_2014}. 

We are pursuing these preliminary results along the following lines:
we will first study the impact of lowering the threshold on the [SII]/H$\alpha$ ratio and $\sigma$ on the number of SNR candidates, which might disfavor SNRs with slow shock velocities and/or older objects \citep{2020MNRAS.491..889K} . We will also include other emission lines in the analysis using the SN1 and SN2 data in hand: for instance, among the SNR candidates in our sample, 47 shows strong [OIII]$\lambda5007$ emission and 27 strong [OII]$\lambda3727$.


\small  
%
\section*{Acknowledgments}   
%
Based on observations obtained with SITELLE, a joint project of Universit\'e Laval, ABB, Universit\'e de Montr\'eal, and the CFHT. We wish to recognize and acknowledge the very significant cultural role that the summit of Mauna Kea has always had within the indigenous Hawaiian community. We are most grateful to have the opportunity to conduct observations from this mountain. 

\bibliographystyle{aj}
\small
\bibliography{proceedings}

\end{document}